# Phase transitions and critical phenomena of tiny grains carbon films synthesized in microwave-based vapor deposition system

**Mubarak Ali** [a, *] **and I –Nan Lin** [b]

[a] Department of Physics, COMSATS University Islamabad, Park Road, Islamabad-45550, Pakistan, *E-mails; mubarak74@mail.com or mubarak74@comsats.edu.pk

[b] Department of Physics, Tamkang University, Tamsui Dist., New Taipei City 25137, Taiwan

**Abstract** –Different peak trends of tiny grains carbon film have been observed under the investigations of Raman spectroscopy and energy loss spectroscopy. Carbon films known in nanocrystalline and ultra-nanocrystalline diamond films are synthesized by employing microwave-based vapor deposition system. Carbon atoms exhibit several state behaviors depending on the incurred positions of their electrons. Different morphology of tiny grains under different chamber pressure is related to different rate of arriving typical energies at/near substrate surface. Those tiny grains of carbon film which evolved in graphitic state atoms are converted to structure of smooth elements where elongation of atoms of one-dimensional arrays is as per exerting surface format forces along opposite poles from their centers. Such tiny grains in the film are the cause of $v_1$ peak under the investigation of the Raman spectrum because of the enhanced propagation of input laser signals through channelized inter-state electron gaps of elongated graphitic state atoms. Those tiny grains of carbon film which evolved in fullerene state are the cause of $v_2$ peak. The tiny grains related to $v_1$ peak possess a low intensity as compared to the ones which comprised atoms having state behaviors known in their exceptional hardness. Tiny grains representing $v_1$ peak in the Raman spectrum are also the cause of field emission characteristic of a carbon film. Different peak recordings were made for the Raman at defined positions indicating a different state of carbon atoms for a different phase of deposited tiny grains, which is in line to their energy loss spectroscopy.

## 1.0 Introduction

Developing materials of selective size and unique application along with investigating their characteristics with precise detail requires a new sort of approach and observation. Phase transition of gaseous state carbon atoms when developing a film in the form of tiny grains is a usual phenomenon. But, it is remained challenging to pinpoint and identify the origin of developing tiny grains where they may introduce the understanding of different features of their film which are not yet discussed and if some of them are discussed, they might be with different explanation as compared to the presented ones. Switching dynamics of morphology and structure of carbon films under their different process conditions have been discussed elsewhere [1].

The syntheses of nanocrystalline diamond (NCD) and ultrananocrystalline diamond (UNCD) films only span over few decades. The interplay of varying process conditions in the presence of different concentrations of feed gases (methane, hydrogen and argon, etc.) are the source of development of different featured films while employing microwave plasma chemical vapor deposition (MPCVD) or microwave plasma enhanced (MPE)-CVD. The characteristics of NCD/UNCD films are mainly recognized in terms of Raman spectroscopy, energy loss spectroscopy and microscopic observations, whereas, their structural performance is mainly based on the field emission characteristics. The NCD/UNCD films mainly contain tiny grains, size ranging tens of nanometer to few nanometers.

It is necessary to understand dynamics of developing tiny-sized particles prior to go for their assembling into large-sized particles [2]. Agglomerations of colloidal matter envisage atoms and molecules to deal them as materials for tomorrow [3]. In certain elements where their atoms execute confined inter-state electron-dynamics, to evolve structure, they involve the conservative forces of relevant formats to engage binding energy [4]. Atoms of suitable elements which execute electron transitions neither ionize, deform nor elongate while inert gas atoms split under the surplus propagation of photons having characteristic of current [5]. The phenomena of heat and photon energy have been discussed where a neutral state silicon atom was considered [6]. Influence of deposition chamber pressure has been discussed in depositing content-specific carbon films while employing hot-filaments deposition system [7]. For the functioning of tiny-sized particles as a diligent application, they may have a pronounced effect to work as either effective nanomedicine or defective because of the certain behaviors of comprised atoms [8].

Electron field emission (EFE) device fabricated by heating tiny diamond grains in hydrogen plasma reveals high performance, which is due to the high defect density and low electron affinity [9]. Increasing nitrogen content slightly increases $sp^2$-bonding in grain-boundary and local bonding structure remains the same [10]. Diverse geometries coupled with varieties of attachment sites enabling diamondoids as useful building blocks for molecular-design applications [11]. Size of grains in UNCD film in the range of 2-3 nm remains thermodynamically stable as tested under broad range of pressure–temperature of hydrogenated surfaces [12]. Grains in nanometer size indicate specific changes because of the decreased mobile dislocations and below their critical size, a softening behavior is often observed, and change is triggered by atomic shuffling where simulation approaches are incapable to evaluate underlying mechanism [13]. Sumant *et al.* [14] investigated the surface chemistry and nanotribology inside the UNCD films and are indistinguishable chemically, adhesively and tribologically from single-crystal diamond.

Several studies of UNCD films are available in the literature targeting field emission characteristics under different process parameters along with varying concentration of gases (and dopants as well). On increasing the local density of states, UNCD films provide the conducting path at diamond-vacuum interface [15]. UNCD films synthesized at low temperatures and in nitrogen environment demonstrated high conductivity in their various applications [16]. Under nitrogen atmosphere, the grain size of UNCD film becomes smaller than 10 nm and metallic conductivity is achieved [17]. UNCD film has low friction/wear and rapid passivation of dangling bonds which is a matter of concern in humid environment [18]. UNCD films are different as compared to NCD films in several ways such as their grain sizes are 2 nm to 5 nm, they are bound by $sp^2$-carbon and usually grow in argon-rich/hydrogen-poor CVD environment and may contain 95–98% $sp^3$-bonded carbon [19]. In UNCD films, homojunction facilitates the tunneling of electrons in conduction band and their direct transport at conducting/insulating interface [20]. To synthesize diamond films at temperatures below 600ºC, UNCD films are superior in choice because their properties don't degrade as in polycrystalline diamond films [21]. On increasing substrate temperature along with hydrogen concentration in Ar/CH$_4$ plasma, $sp^3$-diamond phase becomes less passivated [22]. Low friction coefficient of UNCD films is linked with large amounts of hydroxylic/carboxylic functional groups which are recognized as grain boundary [23]. Bias enhanced nucleation/growth processes facilitate the transport of graphitic phase in both the interior and interface regions of UNCD films [24]. Hydrogen concentration increases the size of diamond grains [25]. Boron-doped UNCD films are capable of being utilized in the field of atomic force microscopy/robust electrodes/bio-compatible sensors [26]. On lowering the a-C layer at interface UNCD/Au-Si, composite structure delivers better conductivity and improves EFE [27]. Precise growth of sub-2 nm/above 5 nm sized diamond grains is achieved due to energetic species under low growth temperature and pre-nucleated interface, thus, tiny grains on silicon nanoneedles deliver enhanced EFE [28]. Gold-UNCD films show superior EFE properties than copper-UNCD films and the authentic factor underneath the mechanism is the presence of nanographitic phases [29]. Copper ion implantation and annealing process increase the EFE due to the improved conducting nature of UNCD films [30]. Presence of graphitic phases at the interfaces of ZnO nanorods/UNCD needles decreases the resistivity of layer and core-shell heterostructures, thus, delivered superior EFE [31]. Nitrogen-incorporated UNCD grains serve as 'high-density diamond nuclei' in the hybrid carbon film [32]. In all those studies and many others, factors related to performance of UNCD/NCD films are linked with the tiny grains of UNCD/NCD films. Improvement in the conductivity but not EFE of P-ions incorporated-UNCD films was due to the transportation of electrons crossing interface [33]. Transportation of electrons is the prime reason for enhanced conductivity and field emission of doped films [34].

In the present work, different 'tiny grains carbon films' are synthesized while employing microwave-based vapor deposition system (MVD) known in MPCVD or MPE-CVD system. Developing process of 'tiny grains carbon films' is discussed under the variation of chamber pressure and different concentration of hydrogen gas. Raman spectroscopy and energy loss spectroscopy of carbon films show different recorded peaks having different intensity and trend. A bit different intensity and trend of peaks resulted when analysis of the same film is made from different locations indicating miscellaneous behaviors of tiny grains in terms of evolving structure. Raman spectroscopy and energy loss spectroscopy analyses of different carbon films tally each other, which are in line to the field emission characteristics. The performance of carbon film in terms of tiny grains elongated atoms of graphitic state is also pinpointed.

## 2.0 Experimental details

To synthesize 'tiny grains carbon films', a microwave-based vapor deposition system (IPLAS-Cyrannus, 2.45 GHz) was employed. A schematic of deposition system is given elsewhere [35]. Silicon wafers N-type nature (ρ ~1-5 Ohm-cm) were used as a substrate. Initial duration of the nucleation process was set 30 minutes, whereas, deposition time was set 60 minutes. Prior to load the samples in the chamber, substrates were cleaned by ultrasonic agitation for the duration of 30 minutes in methanol solution containing nanodiamond powders size ~30 nm and titanium powders of 325 meshes to create nucleation sites. The total mass flow rate was 200 sccm and was kept constant. The substrate temperature was measured by K-type thermocouple where 6 % $H_2$ in argon –rich methane mixture was used to synthesize carbon films and the recorded value was 700°C±10°C, whereas, it was significantly lowered in films synthesized without $H_2$ where the value was 500°C±10°C. While increasing chamber pressure from 100 torr to 200 torr at constant microwave power 1400 (watts), no significant difference in temperature was noted. Due to the lower thermal conductivity of Ar than $H_2$, heating of substrate by plasma is greatly reduced in argon-rich $CH_4$ mixture [16]. Surface morphology of films were analyzed by scanning microscope (ZEISS Model: SIGMA). Peaks related to different phases of tiny grains were identified by Raman spectroscopy analysis (Renishaw; 325 nm and HR800 UV; 632.8 nm). Photon field emission (PFE) known in electron field emission (EFE) of films was measured by using a tunable parallel plate set-up known in $I-V$ curve where an electrometer (Keithley 237) under vacuum level of $10^{-6}$ torr was employed. Dark field, bright field, high-resolution images and selected area photon reflection (SAPR) patterns (known in SAED patterns) of films were taken by the high-resolution microscope (HR-TEM, JEOL JEM2100F) operated at 200 kV. Energy of different bands of films was examined by core loss photon energy loss spectroscopy (core loss PELS) and low loss photon energy loss spectroscopy (low loss PELS), which are known in core loss EELS and low loss EELS, respectively. Details of main parameters for each sample, their analysis and characterization are given in Table 1. Table 1 also shows the direct linkage to analysis and characterization conducted for each sample.

Table 1: Deposition parameters, and analysis and characterization of samples

| Sample # | Chamber pressure (torr) | Ar: $CH_4$: $H_2$ (sccm) | Microwave power (in watts) | Scanning microscope images | Raman spectroscopy analysis | Field emission analysis | BF-, HR-images, SAPR pattern | ELS analysis | Figures |
|---|---|---|---|---|---|---|---|---|---|
| 1 | 100 | 186:2:12 | 1400 | √ | √ | √ | √ | √ | 1, 2, 3, 4, 6, 8, 9, S4 |
| 2 | 150 | 186:2:12 | 1400 | √ | √ | √ | × | × | 1, 2, 3 |
| 3 | 200 | 186:2:12 | 1400 | √ | √ | √ | √ | √ | 1, 2, 3, 5, 7, 8, 9 |
| 4 | 150 | 196:4:0 | 1100 | √ | √ | × | × | × | S1, S2 |
| 5-7 | 100, 150, 200 | 198:2:0 | 1200 | × | × | √ | × | × | S3 (a) |
| 8-10 | 100, 150, 200 | 192:2:6 | 1200 | × | × | √ | × | × | S3 (b) |
| 11-13 | 100, 150, 200 | 186:2:12 | 1200 | × | × | √ | × | × | S3 (c) |

## 3.0 Results

Surface morphology of 'tiny grains carbon films' synthesized at different chamber pressures are shown in Figures 1 (a) to 1 (c); at 150 torr, many tiny grains are in needle-like shapes. In films synthesized at 100 torr and 200 torr, shape of tiny grains is more like round shape. Morphology of 'tiny grains carbon films' alters in microwave-based vapor deposition under different chamber pressures as inductive from the surface topography images shown in Figure 1 (a-c). The different morphology of tiny grains at different chamber pressure is more related to the rate of arriving typical energies near/at substrate surface. This results into evolve tiny grains having different state of comprised carbon atoms. A comprehensive study is discussed [7] presenting the relation between different state behavior of carbon atoms versus the arriving typical energies near/at substrate surface while depositing the carbon films at different chamber pressures in hot-filaments vapor deposition system.

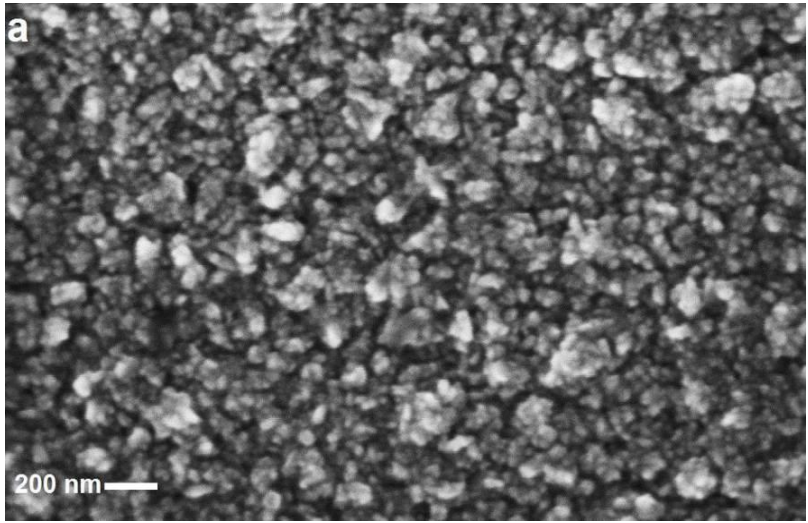

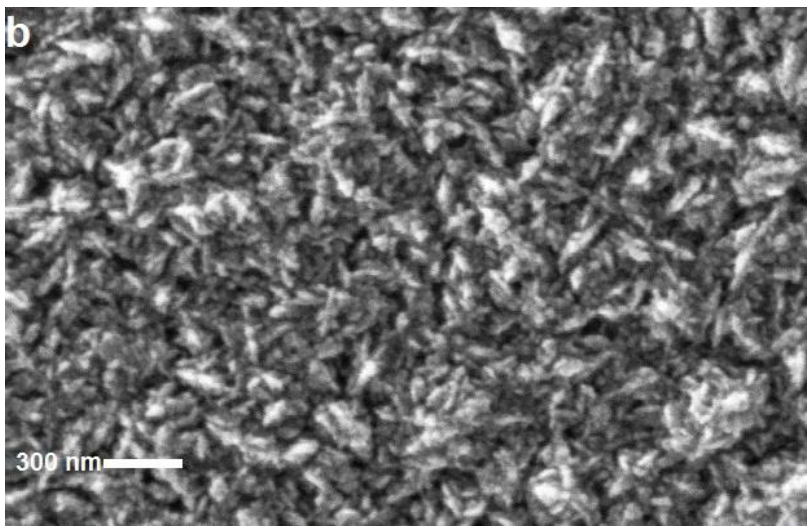

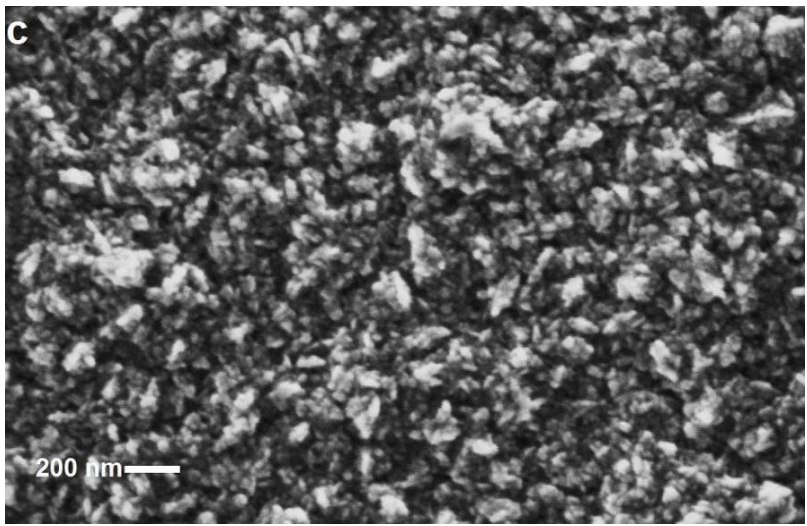

**Figure 1:** Surface morphology of 'tiny grains carbon films' synthesized at chamber pressure (a) 100 torr, (b) 150 torr, and (c) 200 torr; total mass flow rate: 200 sccm (6% $H_2$, 1 % $CH_4$ and 93 % argon, Ar: $CH_4$: $H_2$ = 186:2:12), microwave power: 1400 (in watts) and deposition time: 60 minutes

Raman spectra of different 'tiny grains carbon films' synthesized at 100 torr, 150 torr and 200 torr are shown in Figure 2 where different intensity levels of the different peaks are observed. The $v_1$ peak in different films synthesized at 100 torr and 200 torr is more pronounced as compared to the film synthesized at 150 torr. In different Raman spectra, the $v_1$ peak at wave number ~1150 cm$^{-1}$ is at low level of intensity as compared to peaks related to D*, D and G bands. Raman spectra of different films are also plotted under the in fit multi-peaks as shown in Figure 2 giving the picture at a different angle of the understanding. For different films synthesized at different chamber pressure their $v_1$ peak and $v_2$ peak are also shown in Figure 2 within the limit of varying wave number for their trend.

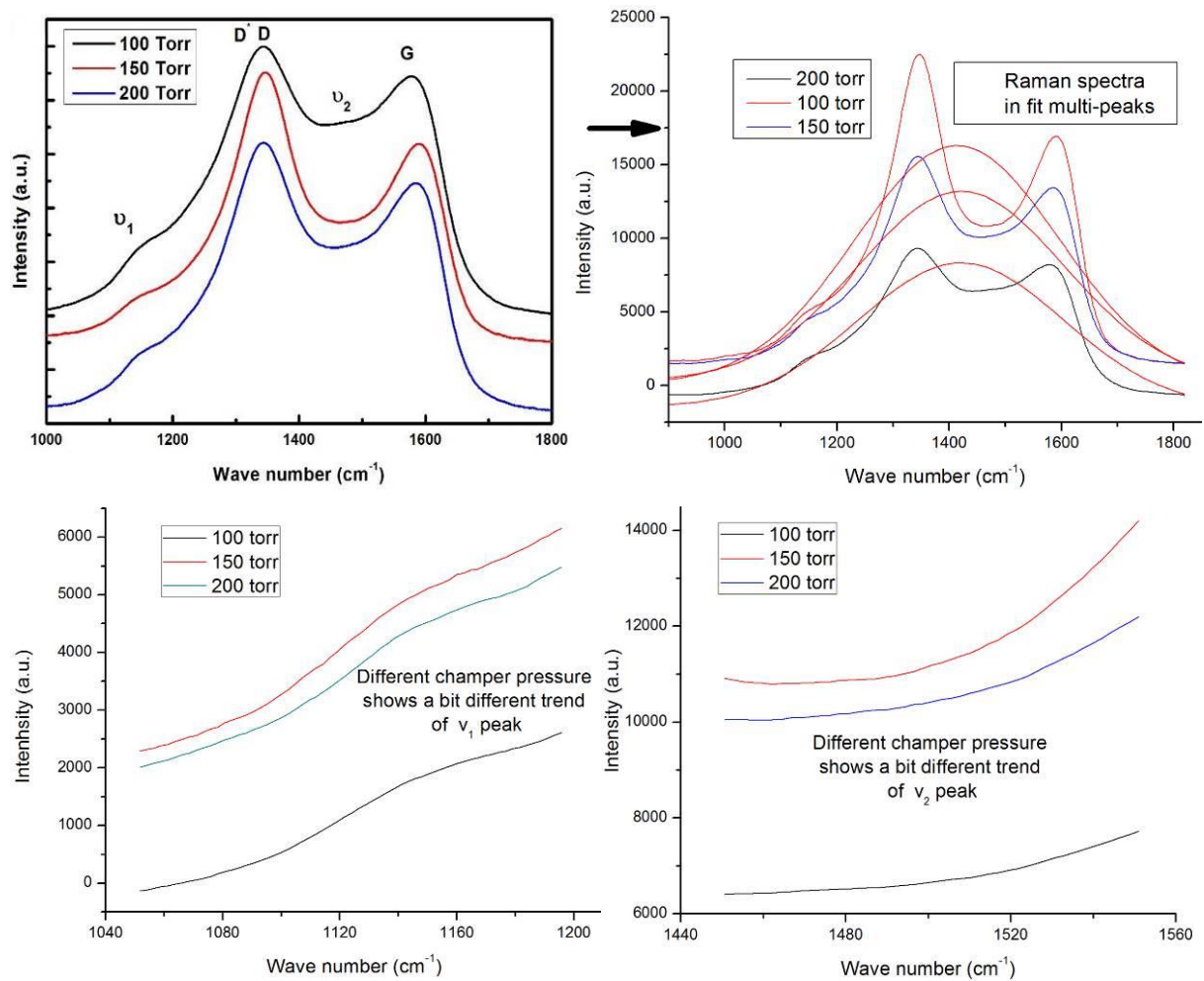

**Figure 2:** Raman spectra of 'tiny grains carbon films' synthesized at chamber pressure 100 torr, 150 torr and 200 torr along with in fit multi-peaks, $v_1$ peak and $v_2$ peak; total mass flow rate: 200 sccm (6% H$_2$, 1 % CH$_4$ and 93 % argon, Ar: CH$_4$: H$_2$ = 186:2:12), microwave power: 1400 (in watts) and deposition time: 60 minutes

A film shown in Figure S1 synthesized without diluting the hydrogen indicates the same morphology of tiny grains as for the case of film deposited under the dilution of hydrogen gas as shown in Figure 1 (b). A pronounced $v_1$ peak is resulted for the film synthesized without diluting hydrogen as shown in Figure S2. However, when hydrogen was incorporated for the initial mixture of gases, the Raman spectrum of the film shows less pronounced $v_1$ peak, which is indicated for three different films synthesized at different chamber pressures as shown in Figure 2 (a-c).

Field emission current density (in known units) as a function of applied field (in known units) at different chamber pressures is shown in Figure 3. Film synthesized at 100 torr shows superior field emission properties where 'turn on field' is 11.52 V/μM (the lowest value) and the large current density (> 2.0 mA/cm$^2$). The 'turn on field' was the highest (25 V/μM) for film synthesized at 150 torr, whereas, it is recorded 16.4 V/μM at 200 torr. Several films synthesized under different amount of hydrogen where only field emission characteristic was studied as presented in the supplementary information (Figure S3). As the deposited carbon films evolved in different morphology of tiny grains (Figure 1), their tiny grains contained different state of carbon atoms. So, in the film having more tiny grains of identical morphology, they evolved in the same state of their carbon atoms. Each carbon state atom undertakes different configuration of the electrons [36]. So, the associated field emission current density will also lead into present different behavior. Therefore, tiny grains carbon films deposited at different chamber pressure while employing the microwave-based vapor deposition will be related to the identical state of carbon atoms where they possess more quantity of tiny grains. Thus, propagating field through tiny grains carbon films deposited at different chamber pressure will measure the different emission current density. Because each state behavior of carbon atoms gives their different allocations of filled and unfilled states of electrons. So, this results into deal the propagation of input field (photons having characteristic of current or different laser signals of different wavelength of photons) in a different manner as well. Current density of applied field is related to density (population) of photons (having characteristic/wavelength of current) giving to a 'tiny grains carbon film' at input end (per unit area or volume), whereas, the field emission current density is related to resulted field from that film at the output end (per unit area or volume). So, the performance of that film (in terms of field emission) depends on the overall structure of its tiny grains.

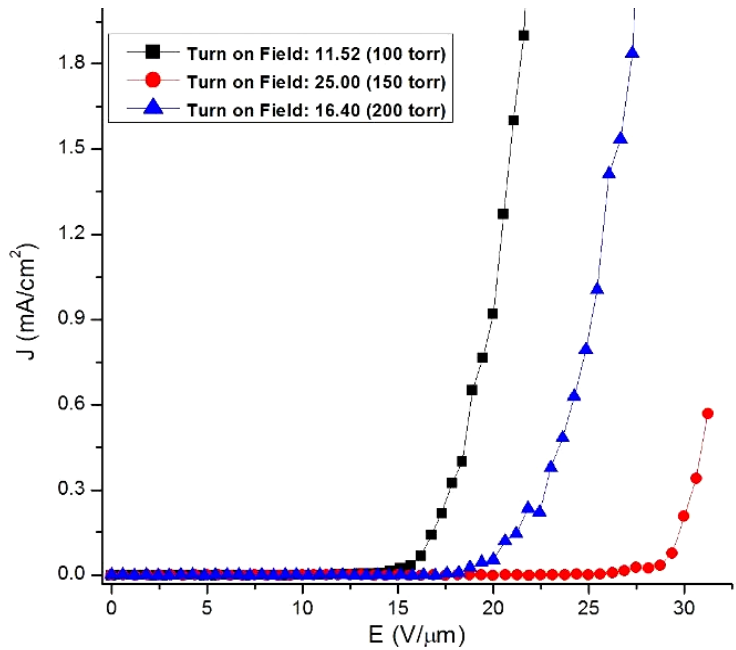

**Figure 3:** Field emission current density of 'tiny grains carbon films' synthesized at chamber pressure 100 torr, 150 torr and 200 torr as a function of applied field (of current dnesity); total mass flow rate: 200 sccm (6% $H_2$, 1 % $CH_4$ and 93 % argon, Ar: $CH_4$: $H_2$ = 186:2:12), microwave power: 1400 (in watts) and deposition time: 60 minutes

Bright field transmission microscope images of 'tiny grains carbon films' synthesized at 100 torr and 200 torr are shown in Figures 4 (a) and 5 (a) where small-sized dark spots are related to the isolated tiny grains and bigger dark spots are related to the coalesced tiny grains. Different shape tiny grains coalesced under the mixed-behavior of exerting forces in the deposition chamber; in some regions of the film, tiny grains amalgamated and in some regions of the film, tiny grains dispersed. SAPR patterns taken at the same regions of films (shown in Figures 4a and 5a) are shown in Figures 4 (b) and 5 (b) where intensity spots don't show the precise rings related to the one phase indicating the presence of different-phase tiny grains. Photons reflected from the surface (of certain orientated-electrons) of atoms of such mixed-dimension structures do not show any specific order in spotted dots of intensity as discernable in their patterns (Figure 4b and Figure 5b).

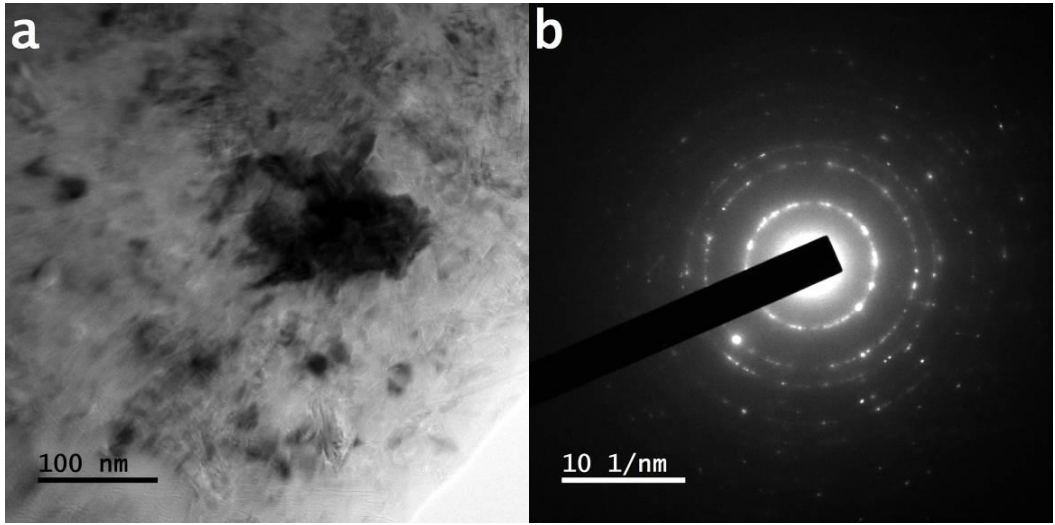

**Figure 4:** (a) Bright field transmission microscope image and (b) SAPR pattern of 'tiny grains carbon film' synthesized at chamber pressure: 100 torr, total mass flow rate: 200 sccm (6% $H_2$, 1 % $CH_4$ and 93 % argon), microwave power: 1400 watts (in watts) and deposition time: 60 minutes

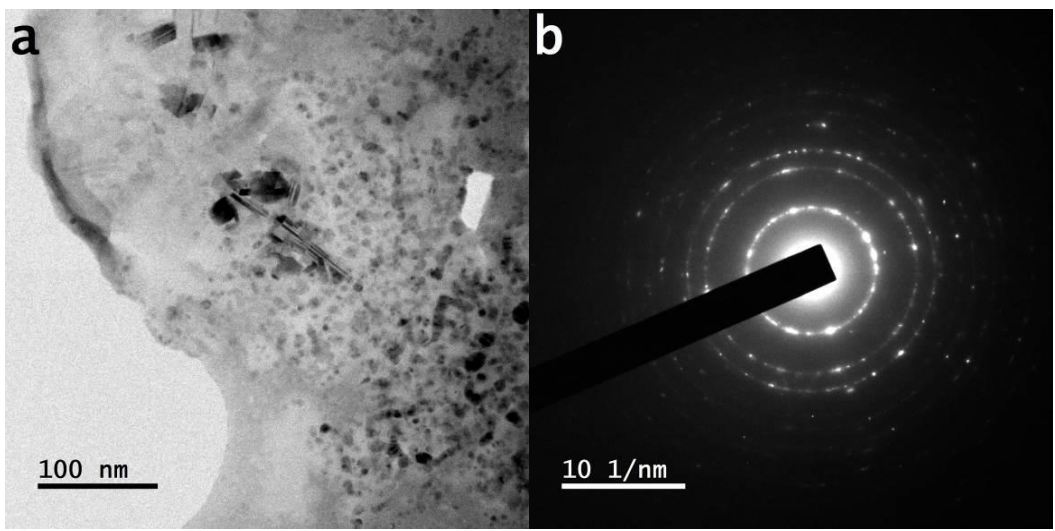

**Figure 5:** (a) Bright field transmission microscope image and (b) SAPR pattern of 'tiny grains carbon film' synthesized at chamber pressure: 200 torr, total mass flow rate: 200 sccm (6% $H_2$, 1 % $CH_4$ and 93 % argon), microwave power: 1400 (in watts) and deposition time: 60 minutes

High-resolution transmission microscope image taken from 'tiny grains carbon film' synthesized at 100 torr is shown in Figure 6 (c) and from the figure; two regions of magnified images are shown in Figures 6 (a) and 6 (b). In Figure 6 (c), several tiny grains show elongation of atoms of one-dimensional arrays where they converted into structure of smooth elements, which are originally related to graphite structure. The elongation of graphitic state atoms is as per stretching of clamping energy knots to electrons along the east-west poles or their near-region poles [36]. In the image shown in Figure 6 (a), atoms of the underlying tiny grain elongated where they converted their one-dimensional arrays into structure of smooth elements prior to deposit the new ones. In Figure 6 (b), a single tiny grain is shown where atoms of one-dimensional arrays elongated uniformly resulting into develop their structure of smooth element. In Figure 6 (c), tiny grains of different textures (surfaces) indicate their different phases in the film where, in more number, they developed in structures of smooth elements as atoms (of graphitic state) having one-dimensional arrays elongated uniformly (along both sides from the centers) under the exertion of surface force at electron-level. Further detail of elongation of atoms is given elsewhere [5].

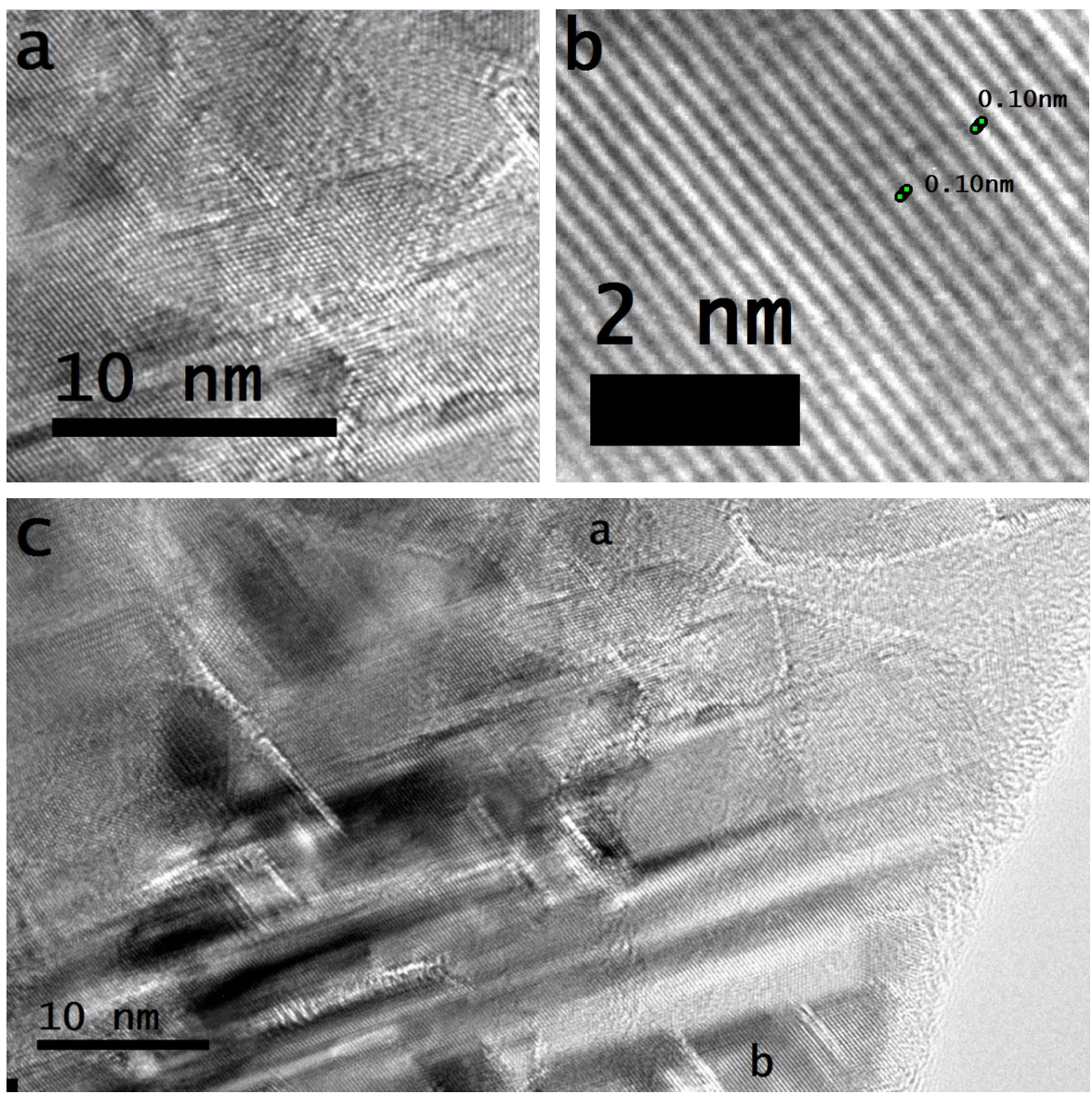

**Figure 6:** (a) Magnified view of the selected region of high-resolution transmission microscope image shows several overlaid tiny grains, (b) magnified view of the selected region of high-resolution transmission microscope image shows elongated atoms of graphite structure tiny grain and (c) standard high-resolution transmission microscope image shows large number of tiny grains developed in structure of smooth elements; 'tiny grains carbon film' synthesized at chamber pressure: 100 torr, mass flow rate: 200 sccm (6% $H_2$, 1 % $CH_4$ and 93 % argon), microwave power: 1400 (in watts) and deposition time: 60 minutes

High-resolution image of tiny grains showing identical features as in Figure 6 (c) was captured from another location of the film, which is shown in Figure S4. The gap between two tiny grains, which were developed in structure of smooth elements, labelled by (2) in Figure S4 indicating canal-like channel and referred to 'graphitic filament' [33, 35]. In the deposition chamber, travelling photons of suitable wavelength further shape structures of smooth elements of such tiny grains by means of aligning states of perturbed (misaligned) electrons of their elongated atoms resulting into provide uniform inter-state electron gap for propagation of photons having characteristic of current.

High-resolution transmission microscope image taken from the film synthesized at 200 torr is shown in Figure 7 (c), which indicates different phases of tiny grains. Two cropped regions of high-resolution image in magnified resolution are shown in Figures 7 (a) and 7 (b). In Figure 7 (a), tiny grains other than those developed in structure of smooth elements are shown. However, in Figure 7 (b), two tiny grains developed in structure of smooth elements are shown where a canal-like channel is formed between them. Different textures of tiny grains are shown in Figure 7 (c), which also indicate different phases where less number of tiny grains developed with structure of smooth elements.

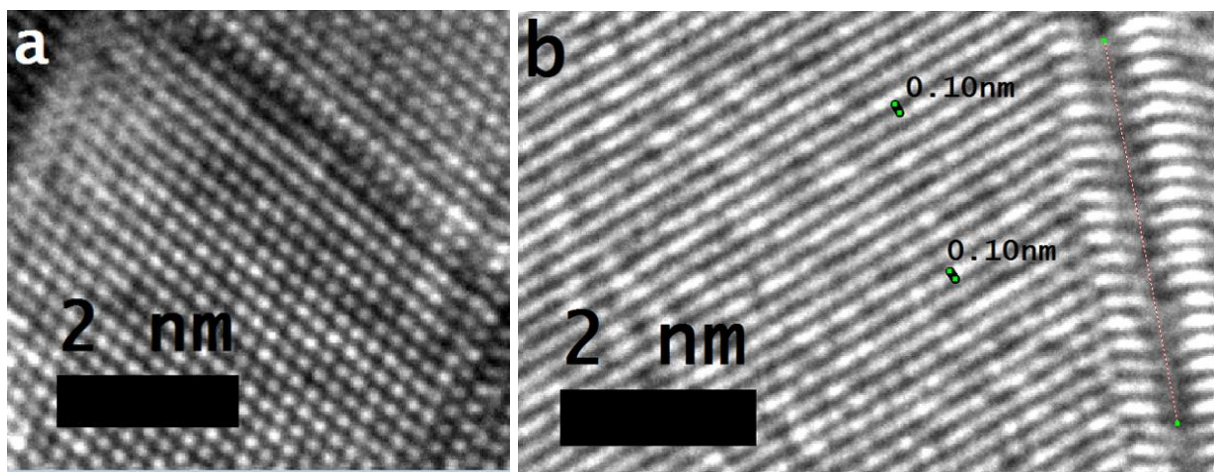

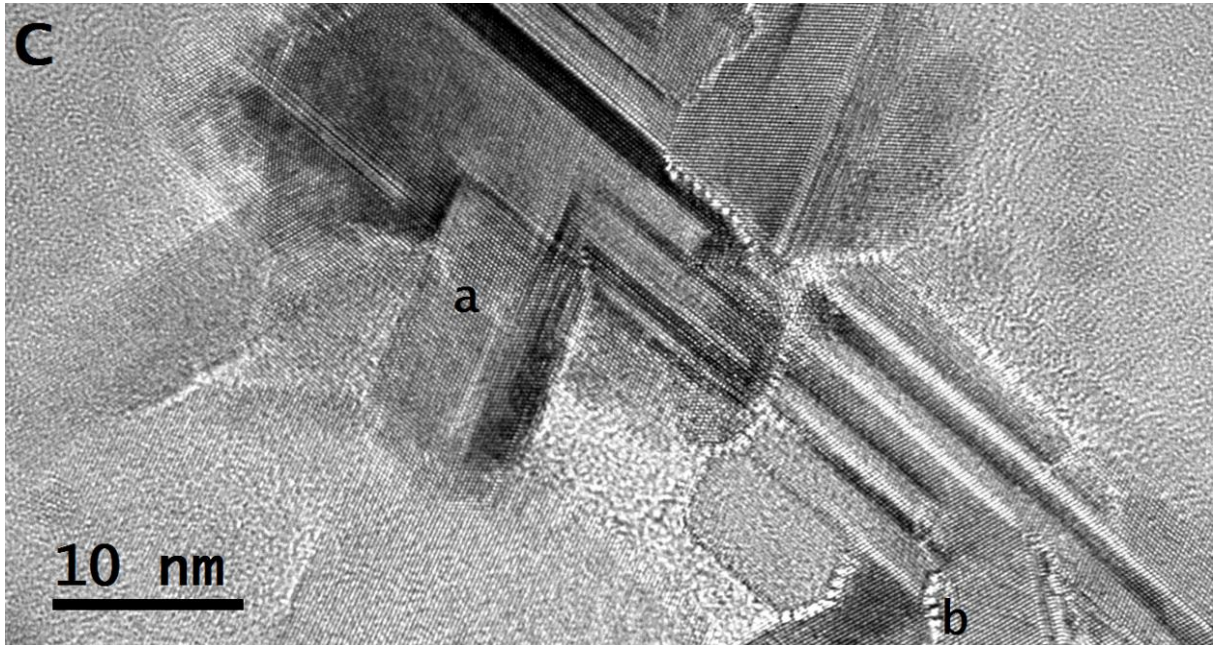

**Figure 7:** (a) Magnified view of the selected region of high-resolution transmission microscope image shows tiny grain of different phases where each phase is related to different state of carbon atoms, (b) magnified view of the selected region of high-resolution transmission microscope image shows tiny grain developed in structure of smooth elements and (c) standard high-resolution transmission microscope image shows different phase tiny grains comprised different state carbon atoms; 'tiny grains carbon film' synthesized at chamber pressure: 200 torr, mass flow rate: 200 sccm (6% $H_2$, 1 % $CH_4$ and 93 % argon), microwave power: 1400 (in watts) and deposition time: 60 minutes

Though grain boundaries are not always clear (blurred) due to resulted resolving power of the microscope because of the not uniformly deposition of those tiny grains in the carbon film, both in the case of Figure 6 (c) and Figure 7 (c), but, it is clear in both images for those tiny grains developed in structures of smooth elements where their atoms were initially in the graphitic state and elongated by the exerting surface force at electron-level (and before the deposition of a new layer of tiny grains over them). Further detail of structure of graphitic state carbon atom along with others is shown in a separate study [36].

From these high-resolution microscopic images, identification of state of carbon atoms for different phases of tiny grains known in their exceptional hardness or other features is not in accordance to the application of microscope. However, it is visible to identify tiny grains evolved in graphitic state carbon atoms as atoms of one-dimensional arrays converted to structures of smooth elements while exerting force at electron-level (in the surface format) as visible from topographic view in all high-resolution transmission microscope images (in Figures 6 and 7). Different state of carbon atoms along with binding of identical state atoms has been discussed elsewhere [36]. In the case where atoms of tiny grains retain graphite structure, they don't elongate due to certain aspects of not exerting forces. They keep maintaining two-dimensional structure of graphitic state atoms once they are hidden from the forces of surface format (or not at the level of oppositely poled exerting forces) because of depositing the new layers of structure together. So, atoms of one-dimensional arrays do not convert to develop their structures of smooth elements. Developing structures of graphitic state atoms in two different ways is discussed in a separate study [36]. In 'tiny grains carbon films', propagation of photons having characteristic of current is accelerated because of the naturally built-in channelized inter-state electron gap of their atoms. This is the cause of recognition of UNCD/NCD films in terms of enhanced field emission characteristics where inter-state electron gaps work efficiently, thus, allowing propagation of accelerated current density of photons [5]. This indicates that atoms of tiny-sized particle don't collectively oscillate on trapping or coupling light (photons), so, disregarding the largely studied phenomenon of surface plasmons or surface plasmon polaritons. However, in tiny grains of graphitic state carbon atoms where they do not develop structure of smooth elements, their atoms may undertake the elongation under the limit of stretching energy knots clamped electrons when they are cleared for the available exerting forces of surface format.

In Figure 8, core loss PELS captured from the same regions of 'tiny grains carbon films' at which high-resolution transmission microscope images were taken. An abrupt rise near 292 eV referred to σ*-band and a large dip near 305 eV referred to tiny grains in diamond state [37, 38] and these bands are also available in films synthesized at 100 torr and 200 torr as shown in Figures 8. However, the large dip curve in both films, which is more indicative in the case of film synthesized at 200 torr (in Figure 8), is due to the energy band related to tiny grains of elongated atoms developed structures of smooth elements (~285 eV). The core loss PELS analysis of both films indicate the same peaks related to different energy bands and similar trend to the one observed in the case of Raman spectroscopy spectrum. But in the case of Raman spectroscopy analysis, the peaks' positions were at different wave numbers, whereas, in core loss PELS analysis, the peaks' positions are at different energy bands; in the carbon film, different phases tiny grains evolved in lonsdaleite, diamond and graphene state atoms originating peaks at position of band energy ~295 (eV), ~308 (eV) and ~328 (eV), respectively. Those tiny grains of carbon film, which are in fullerene state atoms, show the decreasing level of energy in the form of dip curve at energy band ~304 (eV), which is notable in both films as shown in Figures 8. At 100 torr chamber pressure, more pronounced hump-like peaks at ~328 eV are because of a greater number of tiny grains evolved in graphene state atoms.

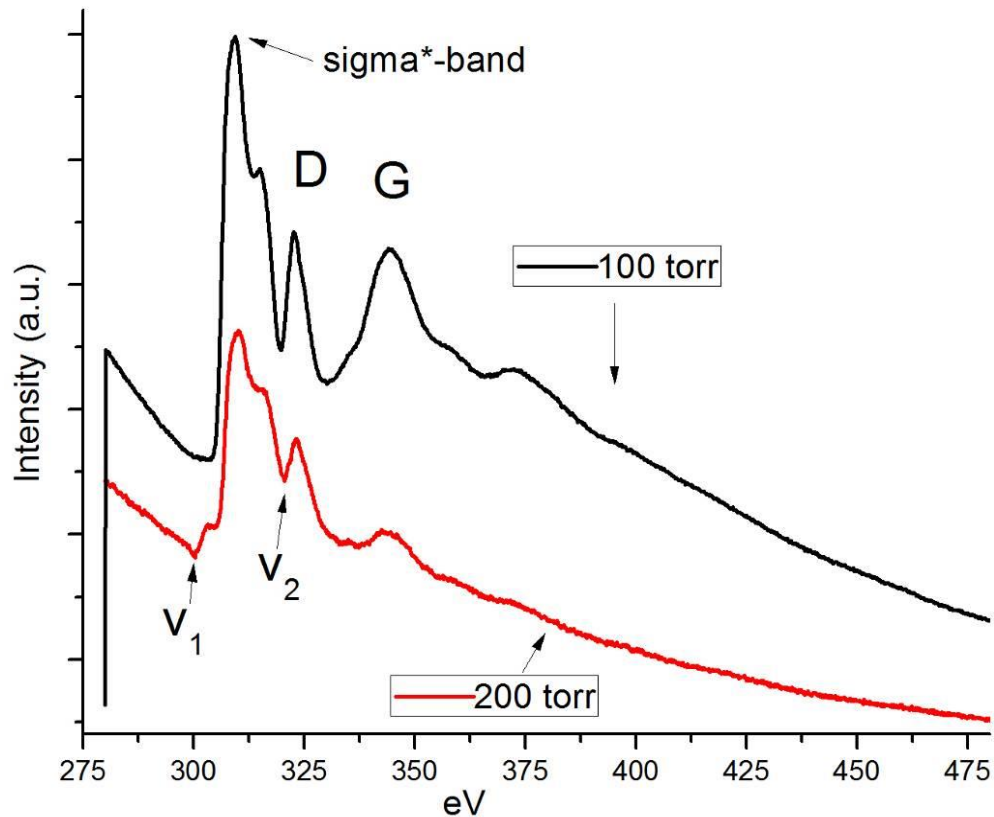

**Figure 8:** Core Loss PELS analysis of 'tiny grains carbon films' synthesized at chamber pressure 100 torr and 200 torr; total mass flow rate: 200 sccm (6% $H_2$, 1 % $CH_4$ and 93 % argon), microwave power: 1400 (in watts) and deposition time: 60 minutes

In low loss EELS analysis, tiny grains of graphitic state atoms show a prominent peak at $s_3$ (27 eV), whereas, the a-C phase shows a peak at $s_1$ (22 eV) [38, 39]. However, in low loss PELS shown in Figure 9, a peak at 27 eV is not available in the analysis and energy band is at slightly titled position instead of to be at 22 eV, which is for both 'tiny grains carbon films' indicating the presence of different phases of tiny grains in the both films. Overall, the trend of peaks' behavior related to different band energy is in the same manner as analyzed in the case of core loss PELS analysis of films. The built-in unit (eV) in the case of core loss PELS analysis and low loss PELS analysis are replaced with photon volts (pV).

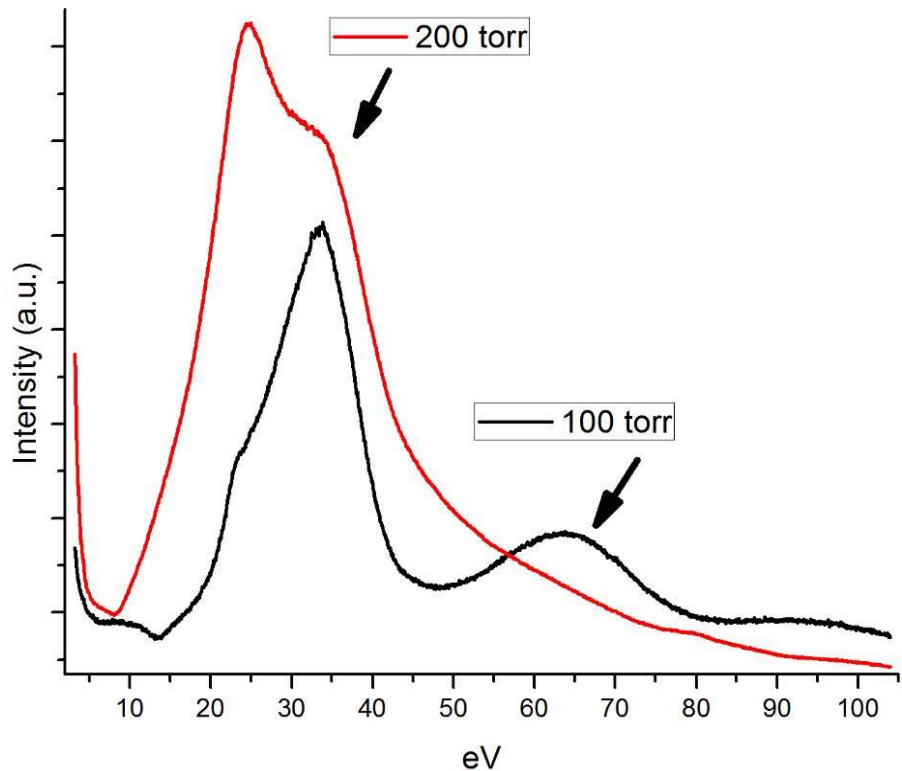

**Figure 9:** Low loss PELS analysis of 'tiny grains carbon films' synthesized at chamber pressure 100 torr and 200 torr; total mass flow rate: 200 sccm (6% $H_2$, 1 % $CH_4$ and 93 % argon), microwave power: 1400 (watts) and deposition time: 60 minutes

**4.0    Discussion**

In Raman spectra of NCD/UNCD films, the peak at 1150 $cm^{-1}$ has been attributed to nanocrystalline diamond [39-41], however, the $v_1$ peak (at 1150 $cm^{-1}$) can't originate from a nanodiamond or related *sp³*-bonded phase but it is assigned to transpolyacetylene segments at grain boundaries and surfaces [42]. Nemanich *et al.* [40, 41] proposed that 1150 $cm^{-1}$ peak arises from nanocrystalline or amorphous diamond. In the Raman spectrum of NCD/UNCD film, the sharp peak at 1332.1 $cm^{-1}$ is the characteristic of diamond and broadening of the peak is related to the NCD grains [43]. In NCD/UNCD films, the peak at 1150 $cm^{-1}$ appears hypothetically due to CH bonds and is always linked to another peak at 1450 $cm^{-1}$, and the peaks at 1330 $cm^{-1}$ and 1560 $cm^{-1}$ are due to D-band and G-band [44], respectively. However, in the light of presented results of 'tiny grains carbon films', argon gas increased the rate of re-nucleation where enabling the development of film in only tiny grains. Depending on how many tiny grains elongated atoms of one-dimensional arrays to develop structures of smooth elements, their quantity is being experienced by the Raman laser, so, the contribution is recorded in the form of peak at wave number ~1150 cm$^{-1}$. In microcrystalline diamond films, the size of grains is very large [1], thus, they don't retain graphite structure and diamond state atoms at a large extent evolving structure under the presence (dilution) of large amount of atomic hydrogen in HFCVD reactor as discussed elsewhere [7], accordingly, information put forth by Raman signals rather to record peak at 1150 cm$^{-1}$, the peak is recorded at 1332 cm$^{-1}$. Thus, the peak at wave number 1150 cm$^{-1}$ is associated to those tiny grains which were developed in the graphite structure where their atoms of one-dimensional arrays undertook elongation while exerting surface format forces resulting into develop the structures of smooth elements. Depending on the amount of tiny grains evolved in graphite structure followed by conversion into structure of smooth elements, the intensity of $v_1$ peak varied; different intensity of $v_1$ peak is related to the number of the tiny grains developed in the structure of smooth elements experiencing the Raman laser. Those tiny grains evolved structure of fullerene state atoms and become the origin of $v_2$ peak in the Raman spectrum. In this way, wherever the $v_1$ peak is resulted in the Raman spectrum of carbon films, the $v_2$ peak is resulted also. Incorporating $H_2$ in Ar-rich $CH_4$ mixture either eliminates the $v_1$ peak or it appears with less intensity. Without incorporation of $H_2$ to Ar-rich $CH_4$ mixture, $v_1$ peak is pronounced as shown in Figure S2. On varying concentration of feed gases, the scale of different phases of tiny grains is altered as well where counting different number of tiny grains developed in structure of smooth elements, thus, recorded different intensity of $v_1$ peak in the Raman spectrum. Again, different intensity of $v_1$ peak in the same film is recorded on dealing Raman signals of different wavelength [45, 46] which is because of experiencing different wavelength signals of employed laser.

In Raman spectroscopy analysis, the peaks at different wave numbers are resulted because of the absorption of energy of interacting Raman laser in a different manner owing to different phases of tiny grains where atoms possessed different states of carbon as shown in Table 2; at wave number ~1145 cm$^{-1}$, the $v_1$ peak related to tiny grains comprised graphitic state atoms where atoms of graphite structure elongated to develop structure of smooth elements; at wave number ~1310

cm$^{-1}$, the peak (due to D* band) related to tiny grains comprised lonsdaleite state atoms; at wave number ~1332 cm$^{-1}$, the peak (due to D band) related to tiny grains comprised diamond state atoms; at wave number ~480 cm$^{-1}$, the v$_2$ peak related to tiny grains comprised fullerene state atom; at wave number ~1580 cm$^{-1}$, the peak related to tiny grains comprised graphene state atoms.

**Table 2:** Identification of different peaks related to different phase tiny grains carbon film under the analyses of Raman spectroscopy and energy loss spectroscopy

| Elongated atoms graphite structure | Lonsdaleite structure | Diamond structure | Fullerene structure | Graphene structure |
|---|---|---|---|---|
| **Raman spectroscopy (in cm$^{-1}$)** | | | | |
| v$_1$ ~1145 (8734 nm) | D* ~1310 (7634 nm) | D ~1332 (7508 nm) | v$_2$ ~1480 (6757 nm) | G ~1580 (6329 nm) |
| **Core loss PELS (in eV)** | | | | |
| Dip curve at ~285 | Peak at ~295 | Peak at ~308 | Dip at ~304 | Peak at ~328 |

In Table 2, wavelength of resulted Raman peak related to different phases of tiny grains available in the deposited film having different state of atoms for each phase is tabulated at different wave number (frequency). The lowest energy of those tiny grains was resulted which had elongated atoms of graphite structure, which is in the form of dip curve at ~285 (in eV) in their core loss PELS analysis, thus, gives the longest recorded wavelength of v$_1$ peak (~8734 nm) in the Raman spectrum analysis. On the other hand, where tiny grains evolved in graphene state atoms, the highest energy of the structure resulted at ~328 (in eV) in their core loss PELS analysis, thus, gives the shortest wavelength (~6329 nm) in the Raman spectrum analysis. Those tiny grains work under the lowest energy and the longest wavelength (shortest wave number) of the structure resulted into the accelerated propagation of photonic current, which is termed as 'elongated atoms graphite structure' and availability of their greater content in film guarantee the enhanced field emission characteristics.

In tiny grains of graphite structure, converted atoms of one-dimensional arrays into structures of smooth elements (elongated atoms graphite structure) further modified to shape their smooth elements if the photons have adequate forcing energy travelling aside to their surface. Thus, such shaped tiny grains perfectly channelized the propagation of photonic current or other input signals under different wavelengths of the laser. A travelling photon is related to forcing energy where energy of quantized amount is forced from one point to another point [6].

As it is tabulated in the Table 2, tiny grains of 'elongated atoms graphite structure' give the highest value of wavelength indicating that they are the cause of enhanced field per unit length or area as discussed above. In the case of tiny grains having state of atoms in lonsdaleite, diamond and graphene, the decreased values of wavelength are resulted in the respective manner. Now the question arises that recorded wavelength in the case of fullerene structure is also between diamond and graphene (in Table 2) but measured hardness is at very low level, however, the peak related to fullerene structure is at very low intensity in the Raman spectrum and we can observe associated energy in the form of dip curve in the spectrum of PELS.

The scheme of electrons is different for different state carbon atoms and electronic structure dedicated to each state carbon atom including graphitic and fullerene is discussed elsewhere [36]. Thus, each structure of established phase tiny grains delivers the different response against coordinated or interacted signals, which is distinctive for each carbon film synthesized at different chamber pressure. Atoms of tiny grains responsible for $v_1$ peak stand independently due to the uniformly specified occupied positions of electrons where their clamping energy knots stretch orientational-based under the exertion of surface format forces. For the case of fullerene state carbon atom, each electron of outer ring occupied each quadrant, therefore, a tiny grain having atoms of fullerene state is responsible to originate $v_2$ peak in the Raman spectrum where path of propagating laser signals is different as compared to tiny grains elongated in graphitic state atoms. So, peaks of Raman spectrum are related to tiny grains of different phases, which are basically as per state behavior of comprised carbon atoms and, so, for the case of energy loss spectrum. Different phases of tiny grains in deposited carbon film are because of the arisen dynamics of atoms amalgamating in few square/cube nanometers where different force-energy behaviors for instantaneous time are involved. The analysis of graphene state atoms presented by many studies and few of them are referred here [47-49]. Similarly, different peaks related to different wave numbers have been discussed elsewhere [50-52] along with in many other studies, however, they originate their different science in the case of present work. The cause of atoms of some elements to be in gas state and some to be in solid state is discussed elsewhere [53]. These findings also solicit to devise science of Raman spectroscopy and energy loss spectroscopy, accordingly.

## 5.0 Conclusions

'Tiny grains carbon films' comprise different phases of tiny grains under different converted state of carbon atoms. A 'tiny grains carbon film' synthesized under suitable conditions while employing microwave-based vapor deposition system involves 'elongated atoms graphite structure' also. In tiny grains where graphitic state atoms bind under the execution of electron-dynamics, they elongated atoms of one-dimensional arrays to structures of smooth elements while exerting the forces of surface format along opposite poles from their centers. Such tiny grains are the cause of $v_1$ peak in the Raman spectrum where the lowest wave number is resulted due to the enhanced field emission performance under the accelerated rate of current density. In the case of tiny grains where carbon atoms converted to fullerene state, they are related to '$v_2$ peak' in the Raman spectrum. Results of field emission measurements of films synthesized at different pressure are in-match to energy loss spectroscopy and Raman spectroscopy analyses. In Raman spectroscopy and energy loss spectroscopy analyses, the peaks related to different wave numbers and energy bands are because of the propagation of input energy signals through inter-state electron gaps of different state carbon atoms evolving tiny grains of different phases in the film.

Traveling photons of adequate wavelength aside to the surface of tiny grains having 'elongated atoms graphite structure' further shape to structure of smooth elements as per their forcing energy, thus, enabling enhanced and accelerated propagation of field under the capacity of different input sources where such tiny grains also negate the phenomenon of surface plasmon polaritons.

Different peaks' trends in Raman spectra and spectroscopic spectra of tiny grains carbon films deposited at different chamber pressures are related to different field emission current density. Different phases of tiny grains in the carbon film occupied elongated graphitic state atoms, lonsdaleite state atoms, diamond state atoms, fullerene state atoms and graphene state atoms where they distinctively determine the nature of their peaks in Raman spectroscopy analysis and energy loss spectroscopy analysis. Presence of tiny grains having different amount of such phases in the film mark different behavior of field emission characteristic and other features.

Our experimental results well prove and justify the position of peaks related to different state carbon atoms evolving tiny grains into different phases of their films synthesized under different conditions of microwave-based vapor deposition system and in good agreement to field emission characteristics. In line with this, present studies set a new horizon in synthesizing, analyzing and evaluating materials at nanoscale, and soliciting reinvestigation of materials' syntheses along with their performances both for existing and future applications.

**Acknowledgements:**

Mubarak Ali thanks National Science Council (now MOST) Taiwan (R.O.C.) for awarding postdoctorship: NSC-102-2811-M-032-008 (2013-2014). Authors wish to thank Mr. Tsung-Min Chen for assisting in synthesis and Dr. Chien-Jui Yeh in TEM operation.

**Authors' biography:**

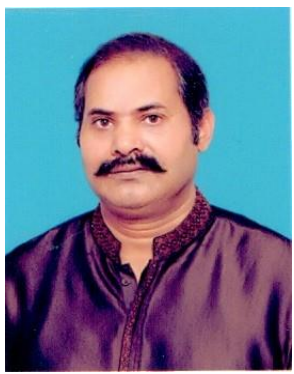

Mubarak Ali graduated from University of the Punjab with B.Sc. (Phys& Maths) in 1996 and M.Sc. Materials Science with distinction at Bahauddin Zakariya University, Multan, Pakistan (1998); thesis work completed at Quaid-i-Azam University Islamabad. He gained Ph.D. in Mechanical Engineering from Universiti Teknologi Malaysia under the award of Malaysian Technical Cooperation Programme (MTCP;2004-07) and postdoc in advanced surface technologies at Istanbul Technical University under the foreign fellowship of The Scientific and Technological Research Council of Turkey (TÜBİTAK; 2010). He completed another postdoc in the field of nanotechnology at Tamkang University Taipei (2013-2014) sponsored by National Science Council now M/o Science and Technology, Taiwan (R.O.C.). Presently, he is working as Assistant Professor on tenure track at COMSATS University Islamabad (previously known as COMSATS Institute of Information Technology), Islamabad, Pakistan (since May 2008) and prior to that worked as assistant director/deputy director at M/o Science & Technology (Pakistan Council of Renewable Energy Technologies, Islamabad; 2000-2008). He was invited by Institute for Materials Research, Tohoku University, Japan to deliver scientific talk. He gave several scientific talks in various countries. His core area of research includes materials science, physics & nanotechnology. He was also offered the merit scholarship for the PhD study by the Government of Pakistan, but he couldn't avail. He is author of several articles published in various journals; https://scholar.google.com.pk/citations?hl=en&user=UYjvhDwAAAAJ, https://www.researchgate.net/profile/Mubarak_Ali5.

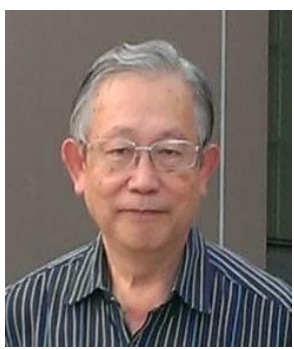

I-Nan Lin is a senior professor at Tamkang University, Taiwan. He received the Bachelor degree in physics from National Taiwan Normal University, Taiwan, M.S. from National Tsing-Hua University, Taiwan, and the Ph.D. degree in Materials Science from U. C. Berkeley in 1979, U.S.A. He worked as senior researcher in Materials Science Centre in Tsing-Hua University for several years and now is faculty in Department of Physics, Tamkang University. Professor Lin has more than 200 referred journal publications and holds top position in his university in terms of research productivity. Professor I-Nan Lin supervised several PhD and Postdoc candidates around the world. He is involved in research on the development of high conductivity diamond films and on the transmission microscopy of materials.

## Supplementary Information:

To further support the investigations of results and discussion and to go into broader insight of $v_1$ peak in the Raman spectrum, results of another 'tiny grain carbon film' deposited at different process parameters are considered for which the morphology of film is shown in Figure S1. The Raman spectrum of the film is shown in Figure S2. The Raman spectrum of film shown in Figure S2 shows the $v_1$ peak more pronounced as compared to the Raman spectra of films synthesized at 6% $H_2$ (in Figure 2a-c); the film was synthesized without incorporating hydrogen gas in the mixture which was the main cause of originating the pronounced $v_1$ peak. As the Raman spectrum shown in Figure S2 is without the dilution of hydrogen gas, so, this indicates the sensitivity of phase transition of tiny grains with a large difference. This requires further investigations under different process parameters for the processing of 'tiny grains carbon films' in microwave-based vapor deposition process. The surface topography view in Figure S1 shows morphology of tiny grains of more like the morphology of tiny grains of film synthesized at 6 % $H_2$ (Figure 1b); chamber pressure and microwave power were not kept the same as prescribed in their Figure's caption and, thus, might introduce the role to evolve films of tiny grains in identical features morphology.

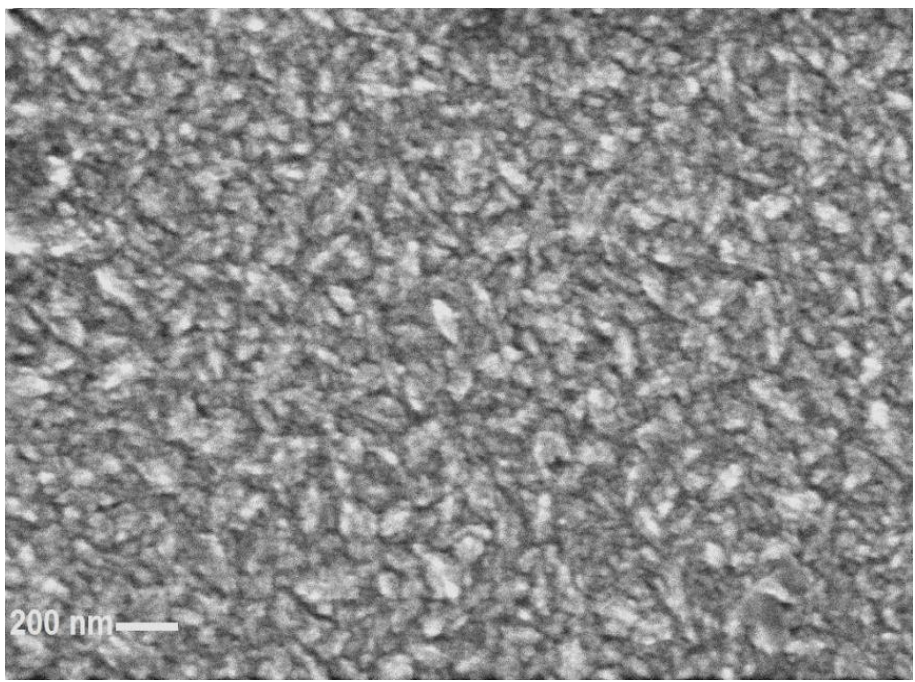

**Figure S1:** Surface morphology of 'tiny grains carbon film' synthesized at chamber pressure 150 torr; total mass flow rate: 200 sccm (0% $H_2$, 2 % $CH_4$ and 98 % argon, Ar: $CH_4$: $H_2$ = 196:4:0), microwave power: 1100 (in watts) and deposition time: 60 minutes

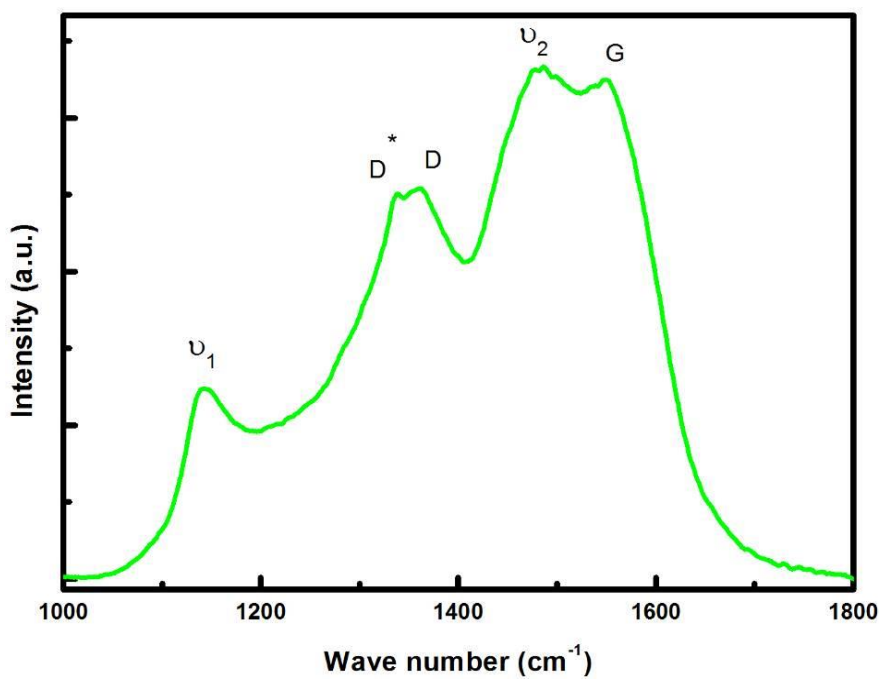

**Figure S2:** Raman spectrum of 'tiny grains carbon film' synthesized at chamber pressure 150 torr; total mass flow rate: 200 sccm (0% $H_2$, 2 % $CH_4$ and 98 % argon, Ar: $CH_4$: $H_2$ = 196:4:0), microwave power: 1100 (in watts) and deposition time: 60 minutes ($v_1$: ~1145 $cm^{-1}$, D*: ~1310 $cm^{-1}$, D: ~1332 $cm^{-1}$, $v_2$: ~1480 $cm^{-1}$, G: ~1580 $cm^{-1}$)

To further validate our investigations, more 'tiny grain carbon films' were synthesized under different process parameters and only their field emission characteristics were performed. Field emission current density of different 'tiny grains carbon films' (synthesized for Ar-rich $CH_4$ mixture without $H_2$ and with 3% and 6% $H_2$) versus applied field is shown in Figures S3 (a-c). In Figures S3 (a) to S3 (c), the general trend of the 'turn on field' is more morphology-dependent; the value of 'turn on field' at 100 torr is lower than that at 150 torr, which is in line to the results of field emission shown in Figure 3. However, considered films were synthesized under different microwave power (1200 watts) as compared to the ones shown in Figure 3. Under lower chamber pressure (at 100 torr), all films in Figure S3 (a-c) show lower 'turn on field' indicating more tiny grains of deposited film elongated their atoms of one-dimensional arrays to develop structures of smooth elements as lower 'turn on field' is resulted (measured) for their analysis. Therefore, the large value of 'turn on field' in film synthesized at 200 torr (Figure S3a) indicates a smaller number of tiny grains which converted atoms of one-dimensional arrays into structures of smooth elements. The results presented in Figure S3 don't contain data for other analyses and characterizations where the presented analysis is based on extracted information from the results and published studies for field emission applications. At different amounts of hydrogen, the rate of their trapping inside the tiny grains carbon films along with association to different morphology tiny grains eventually alter resulting into determine (measure) different rate of field emission current density.

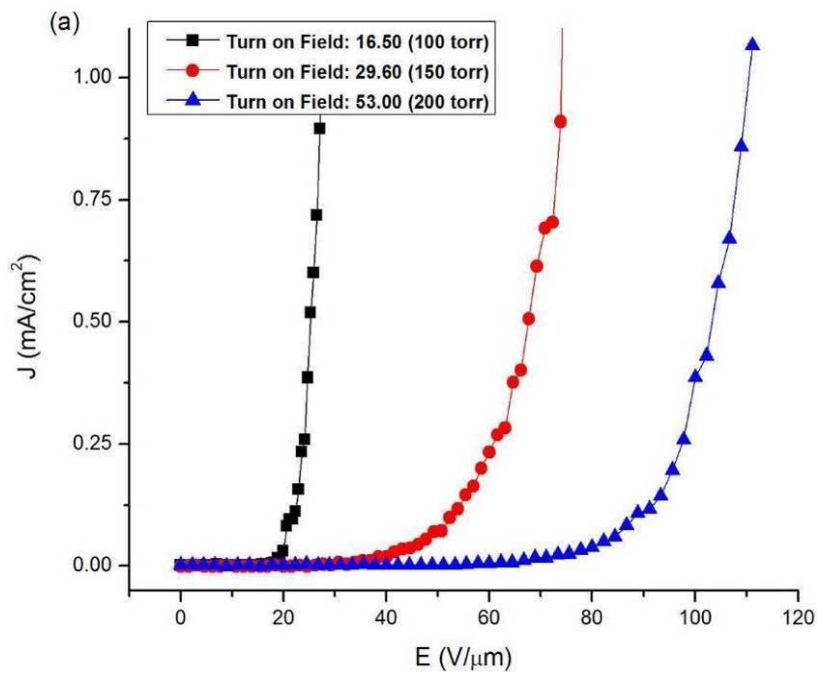

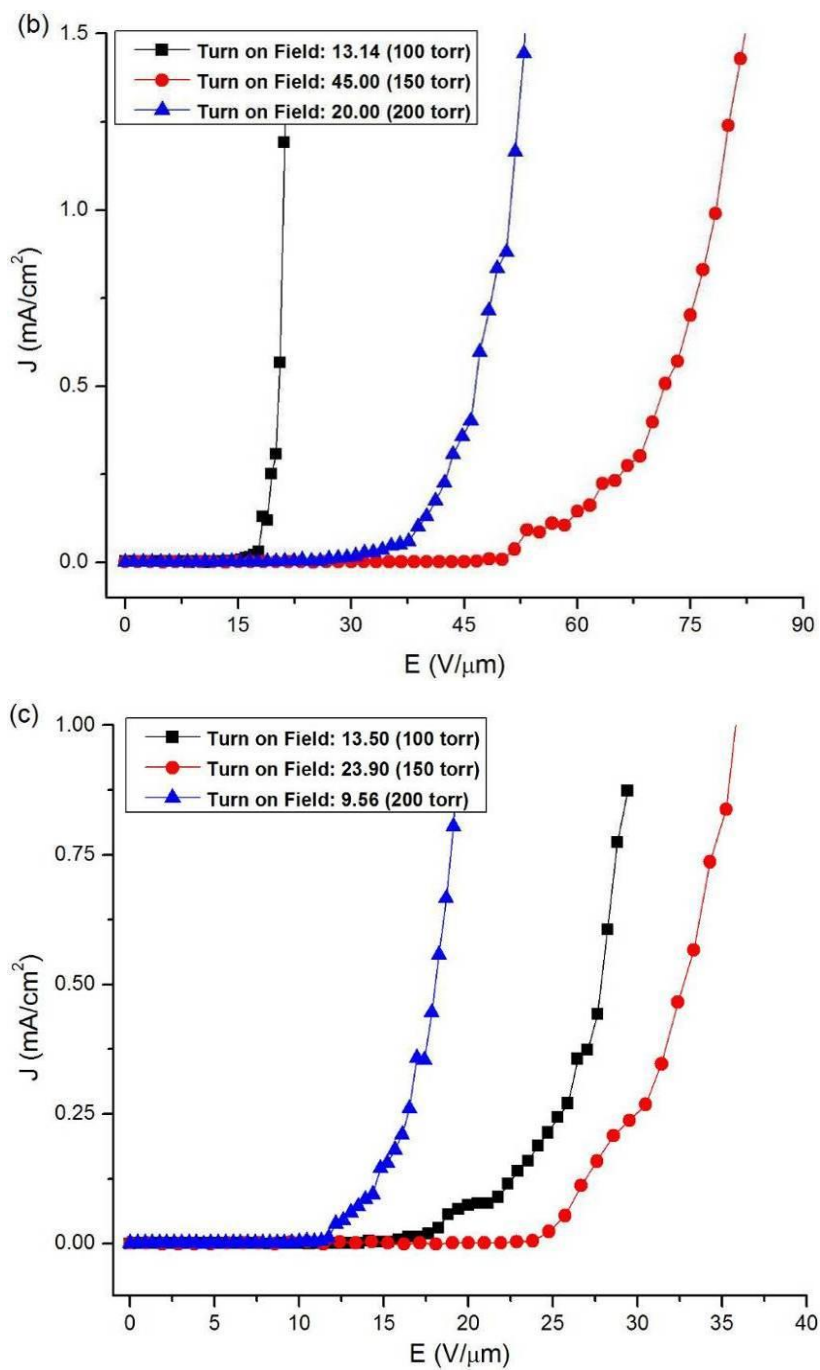

**Figure S3:** Field emission current density versus applied field of 'tiny grains carbon films' synthesized (a) without $H_2$, (b) with 3 % $H_2$ and (c) with 6 % $H_2$; 1 % $CH_4$, total mass flow rate: 200 sccm, microwave power: 1200 (in watts) and deposition time: 60 minutes

Figure S4 shows high-resolution transmission microscope image taken from another region of the 'tiny grains carbon film' synthesized at 100 torr champer pressure highlighting different features of the tiny grains.

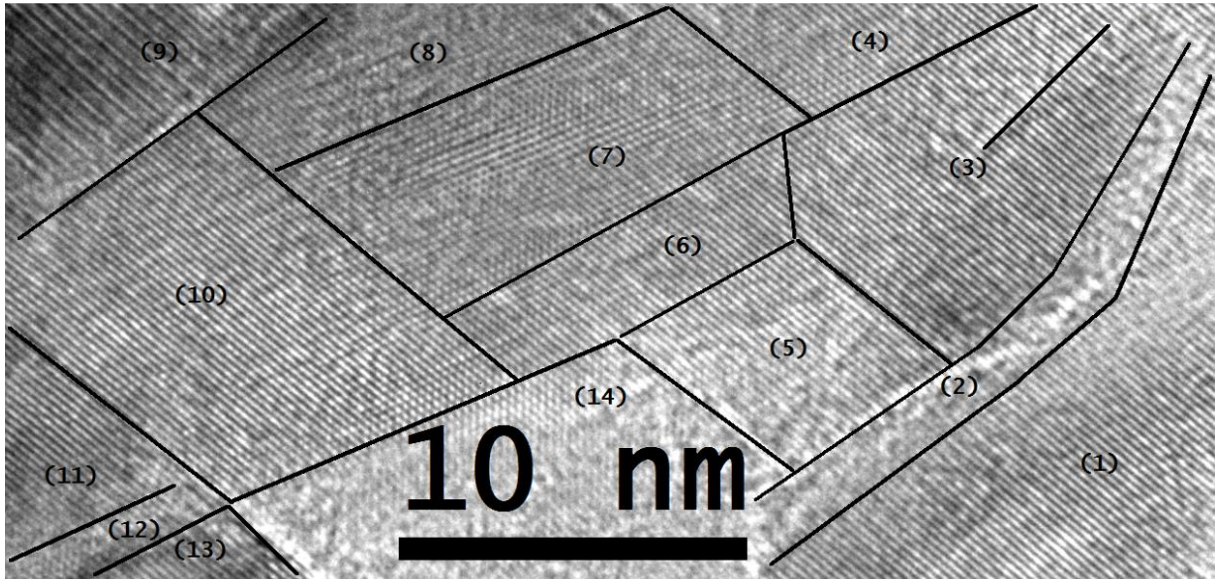

**Figure S4:** High-resolution transmission microscope image of carbon film synthesized at chamber pressure 100 torr shows different phases of tiny grains; total mass flow rate: 200 sccm (6% $H_2$, 1 % $CH_4$ and 93 % argon), microwave power: 1400 (in watts) and deposition time: 60 minutes